\documentstyle [12pt]{article}
\begin{document}

\begin{titlepage}
%\rightline{DRAFT}

\rightline{ULB-PMIF/93-03}
\vspace{1cm}
%\vspace{3cm}
\begin{centering}

{\Large \bf Remarks on the renormalization of gauge invariant
operators in Yang-Mills
theory}

\vspace{2cm}

{\large  Marc Henneaux}\\
\vspace{1cm}
Facult\'e des Sciences, Universit\'e Libre de Bruxelles,\\
Campus Plaine C.P. 231, B-1050 Bruxelles, Belgium\\
and\\
 Centro de Estudios Cient\'{\i}ficos de Santiago,\\
 Casilla 16443, Santiago 9, Chile\\

\vspace{2.5cm}
{\large Abstract}
\vspace{.5cm}

\end{centering}

A simplified proof of a theorem by Joglekar and Lee on the
renormalization
of local gauge invariant operators in Yang-Mills theory is given.
It is based on
(i) general properties of the antifield-antibracket formalism;
and
(ii) well-established results on the cohomology of
semi-simple Lie
algebras.

\end{titlepage}

\thispagestyle{empty}
\vfill
\pagebreak

\section{Introduction}

The BRST transformation in Yang-Mills theory was originally
defined without
reference to the equations of motion and incorporated only
the gauge
symmetry \cite{BRS,Tyutin}.  However, it was subsequently
found necessary to
take the equations of motion into account.  This
can be achieved by
introducing one ``antifield"
for each field appearing in the path integral, and by
defining the BRST variation of the antifields in
such a way that the BRST
differential implements the equations of motion
$D_\mu F^{\mu \nu}_a = 0$
in cohomology.  These
developments were first pursued in the context of
the renormalization of the
Yang-Mills field \cite{ZJ}. They then turned out to
be crucial for the
BRST formulation of arbitrary gauge theories with
``open" algebras \cite{BV},
for which one cannot separate the gauge symmetry
from the dynamics.  The key
role played by the equations of motion and the
concept of covariant phase
space \cite{Kijowski,Witten,Ashtekar,Zuckerman} in
the antifield formalism was
emphasized in J\cite{Fisch,Henn1,HT} (see also
\cite{Alfaro} and references
therein for a related but different point of view).

In the case of the pure Yang-Mills field, the
complete BRST transformation
reads, in the minimal sector containing the vector potential
$A^a_{\mu}$, the ghosts $C^a$, and their
respective antifields $A^{*\mu}_a$,
$C^*_a$,
\begin{equation}
s = \delta + \sigma,
\end{equation}
where $\delta$ is defined by
\begin{equation}
\delta A^a_\mu = 0,\;\; \delta C^a = 0,\; \;
\delta A^{*\mu}_a = D_{\nu} F^{\nu \mu}_a,\;\;
 \delta C^*_a = D_{\mu}A^{*\mu}_a,
\end{equation}
and $\sigma$ is the original BRST differential of
\cite{BRS,Tyutin},
\begin{equation}
\sigma A^a_\mu = D_{\mu} C^a,\;\; \sigma C^a =
{\frac{1}{2}} C^a_{bc} C^b C^c,
\end{equation}
extended to the antifields so as to anticommute with $\delta$,
\begin{equation}
\sigma A^{*\mu}_a = - C^b_{ac} A^{*\mu}_b C^c,
\;\; \sigma C^*_a = - C^b_{ac} C^*_b C^c.
\end{equation}
Here, $C^a_{bc}$ are the structure constants of the
Lie algebra ${\cal G}$ of
the gauge group, which we assume to be semi-simple.  One has
\begin{equation}
\delta^2 = 0, \;\; \sigma^2 = 0, \;\; \delta \sigma +
 \sigma \delta = 0,
\end{equation}
so that
\begin{equation}
s^2 = 0.
\end{equation}
In order to fix the gauge, one introduces also
the antighosts $\overline{C}_a$ and the auxiliary fields
$b_a$, together with their respective antifields
$\overline{C}^{*a}$ and $b^{*a}$.
In this ``non minimal" sector, the BRST differential
is defined by
\begin{equation}
s \overline{C}_a = b_a, \;\; s b_a = 0, \;\;
 s \overline{C}^{*a} = 0, \;\;
s b^{*a} = - \overline{C}^{*a}.
\end{equation}
The fields and antifields have ghost number given by
\begin{equation}
gh A^a_\mu = 0, \;\; gh C^a = 1, \;\; gh A^{*\mu}_a = - 1, \;\;
gh C^*_a = - 2,
\end{equation}
\begin{equation}
gh \overline{C}_a = - 1, \;\; gh b_a = 0, \;\;
gh \overline{C}^{*a} = 0, \;\; gh b^{*a} = - 1.
\end{equation}
We shall denote by ${\cal A}$ the algebra of
 polynomials in $A^a_\mu$,
$C^a$, $A^{*\mu}_a$, $C^*_a$, $\overline{C}_a$,
$b_a$, $\overline{C}^{*a}$,
$b^{*a}$ and their derivatives up to a finite order;
 and by ${\cal B}$ the
smaller algebra of polynomials in $A^a_\mu$, $C^a$,
$A^{*\mu}_a$, $C^*_a$ and
their derivatives up to a finite order.

The nilpotency of $s$ enables one to define
cohomological groups
$H^*(s)$ in the standard manner.  The following
question concerning the
cohomology of $s$ in ${\cal A}$ is of central
interest in the analysis of the
renormalization of local gauge invariant
operators
\cite{Georgi,Gross,Sarkar,Dixon,KS,Joglekar,Deans,Voronov,Collins}: given
a polynomial in  ${\cal A}$ that
is (i) BRST-invariant and (ii) of ghost number zero,
can it be written
as the sum of a gauge invariant polynomial and a
BRST variation?  In other words,
does one have
\begin{equation}
s P = 0 ,\; gh P = 0,\; P \in {\cal A}
\Longrightarrow  P =  M + s Q
\end{equation}
where $M$ is a gauge invariant polynomial in the field
strengths $F^a_{\mu \nu}$
and their derivatives up to a finite order,
and where $Q \in {\cal A}$?  Differently put,
can one find in each
equivalence class of $H^0(s)$ a representative
that does not involve the
ghosts and that is strictly gauge invariant?
The answer to this
question was conjectured inJ \cite{KS}
to be the affirmative.  A proof of (10) has been
given in \cite{Joglekar}
following earlier work of \cite{Dixon}.  The purpose
 of this letter is to
provide an alternative and somewhat shorter proof,
which is based on known results on the antifield
formalism and the cohomology
of Lie algebras.

\section{Digression}

The question (10) was actually formulated originally
in terms of the gauge-fixed action.  In order
to make contact with
the original formulation,
let us denote collectively all the fields
$A^a_\mu, C^a, \overline{C}_a$
and $b_a$ by $\phi^A$ and all the antifields
$A^{*\mu}_a, C^*_a,
\overline{C}^{*a}$ and $b^{*a}$ by $\phi^*_A$.
The algebra of polynomials in
the fields $\phi^A$ and their derivatives up to
a finite order is denoted by
${\cal F}$.  The gauge fixed action is the
integral of an element of
${\cal F}$ obtained by eliminating the
antifields $\phi^*_A$ through
$\phi^*_A = \delta \Psi / \delta \phi^A$,
where $\Psi$ is the so-called
gauge fixing fermion \cite{BV}.

The gauge-fixed BRST symmetry $s_\Psi$
is defined in the algebra
${\cal F}$ through $s_{\Psi} \phi^A =
s \phi^A (\phi^B, \phi^*_B =
\delta \Psi / \delta \phi^B)$ and is a symmetry
of the gauge-fixed action. In
the case of the Yang-Mills theory considered
 here, $s_\Psi$ does not depend on the
gauge fixing since $s \phi^A$ does not depend
on the antifields, $s_\Psi \phi^A =
s \phi^A$.  Thus, $s_\Psi$
is nilpotent off-shell (although, in general,
$s_\Psi$ is only nilpotent
on-shell).  The gauge-fixed BRST cohomology
$H^*_\Psi(s)$ is defined
to be the
set of equivalence classes of weakly BRST
invariant elements of ${\cal F}$,
\begin{equation}
s R \approx 0, \; R \in {\cal F}
\end{equation}
modulo weakly BRST exact ones,
\begin{equation}
R \sim R' \;\hbox{ iff } \; R \approx R' + s T,
 \;\;\; R, R', T \in {\cal F}.
\end{equation}
One says that two polynomials in ${\cal F}$ are
``weakly equal" (for the gauge
conditions under consideration) if they
coincide when the equations of motion following from
the gauge-fixed action hold. They then differ by
a combination of
these equations of motion and their spacetime
derivatives.

As shown in
\cite{Henn1,HT}, the gauge-fixed BRST
cohomology $H^*_\Psi(s)$ defined in the
algebra ${\cal F}$ of the fields is
isomorphic with the BRST cohomology $H^*(s)$
defined in the algebra ${\cal A}$
of the fields and the antifields.
Furthermore, if $P \in {\cal A}$ defines an
element of $H^*(s)$, then
$R = P(\phi, \phi^* = \delta \Psi / \delta \phi) \in {\cal F}$
defines the
corresponding element of the gauge-fixed BRST cohomology
$H^*_\Psi(s)$.
The question
(10) is thus equivalent to : does
\begin{equation}
s R \approx 0, \;\; R \in {\cal F}
\end{equation}
imply
\begin{equation}
R \approx M + s T,\;\; T \in {\cal F}
\end{equation}
where $M$ is a gauge-invariant polynomial in the
field strengths and
their derivatives up to a finite order?  In some
gauges (``linear
gauges"), Eqs. (13) and (14) can
be further simplified because the equation of
motion of the antighost can
be easily handled.  However, even though
$H^*_\Psi(s)$ refers
explicitly to a definite gauge fixation through
the gauge-fixed equations
of motion used in Eq. (12), it  does not
depend on the choice of gauge since it is
isomorphic to $H^*(s)$.
Hence, its
calculation can be carried out independently  of
the gauge fixation.  This
leads to the formulation (10) in terms of the
antifields.  The gauge
independence of the cohomological questions behind
renormalization theory
has been particularly stressed in
\cite{Voronov,Tyutin2}.  After this brief
digression, we can return to the proof of Eq. (10).

\section{Removing the non-minimal sector}

The first step in the proof of Eq. (10)
consists in removing the non-minimal
sector.

\noindent
{\bf Theorem 1} : {\it In each equivalence
class of $H^*(s)$, one can find a
representative that does not depend on the
variables of the non
minimal sector.  That is, if $sP = 0, P \in {\cal A}$, then
$P = R + s Q'$ with
$R \in {\cal B}, s R = 0, Q' \in {\cal A}$.}

\noindent
{\bf Proof} : See \cite{Fisch,Henn1,HT}
(the BRST transformation in
the non-minimal sector takes the characteristic
contractible form
$sx_i = y_i, sy_i = 0$) and \cite{Henn2} for
the implementation of
locality.

We shall thus assume from now on that $P$ in Eq. (10)
belongs to the polynomial
algebra ${\cal B}$ generated by the variables of
the minimal sector and their
derivatives.  In that
algebra, we introduce a second grading, called
the ``antighost number" through
\begin{equation}
antigh A^a_\mu = 0, \; antigh C^a = 0, \; antigh A^{*\mu}_a = 1, \;
antigh C^*_a = 2, \; antigh \partial_\mu = 0.
\end{equation}
The splitting (1) of the differential $s$
simply corresponds to the splitting
according to definite antighost number,
\begin{equation}
antigh \delta = - 1, \;\; antigh \sigma = 0.
\end{equation}
As shown in \cite{Fisch,Henn1,HT}, the
differential $\delta$ is the
Koszul-Tate differential associated with
the gauge-invariant equations of
motion $D_\mu F^{aJ\mu \nu} = 0$.  One has
\begin{equation}
H_k (\delta) = 0, \;\; \hbox{ for } k \not= 0.
\end{equation}
both in the algebra of all functionals
\cite{Fisch,Henn1,HT} and
in the algebra ${\cal B}$ of local polynomials
\cite{Henn2}, while $H_0(\delta)$
is given by the equivalence classes of
polynomials in ${\cal B}$ that coincide
when $D_\mu F^{aJ\mu \nu} = 0$.

\section{Cohomology of $\sigma$}

The cohomology of $\sigma$ - the usual BRST
differential - has
been computed by various authors
\cite{Dixon2,Band,Brandt,DubV} in the polynomial
algebra
generated by the potentials $A^a_\mu$,
the ghosts $C^a$ and their
derivatives.  One can easily include the
antifields $A^{*\mu}_a$,
$C^*_a$ and their derivatives as follows.
The algebra ${\cal B}$ is the sum of
two algebras, ${\cal B} = {\cal C} \oplus {\cal D}$,
one of which, namely
${\cal C}$, is contractible.  The algebra
${\cal C}$ is the
polynomial algebra in the components $A^a_\mu$
of the vector potential, their
symmetrized derivatives $\partial_{(\mu_1 \mu_2 ...
\mu_k} A^a_{\mu)}$ and their $\sigma$-variations \cite{DubV}.
It is contractible because it
takes the standard form $\sigma x_i = y_i, \sigma y_i = 0$.
The algebra ${\cal D}$ is the polynomial algebra
generated by the components
of the field strengths $F_{\mu \nu}^a$, their
covariant derivatives
$D_{\mu_1} D_{\mu_2}...D_{\mu_k} F^a_{\mu \nu}$,
the components of the
antifields $A^{*J\mu}_a$, their covariant derivatives
$D_{\mu_1} D_{\mu_2} ... D_{\mu_k} A^{*J\mu}_a$,
the antifields $C^*_a$, their
covariant derivatives $D_{\mu_1} D_{\mu_2} ... D_{\mu_k} C^*_a$
and the ghosts $C^a$ (without their derivatives,
which are in ${\cal C}$).
One has
\begin{equation}
\sigma \chi^a_\Delta = C^a_{bc} \chi^b_\Delta C^c
\end{equation}
\begin{equation}
\sigma C^a = \frac {1}{2} C^a_{bc} C^b C^c
\end{equation}
where $\chi^a_\Delta$ stands for
$D_{\mu_1 \mu_2 ...\mu_k} F^a_{\mu \nu}$,
$D_{\mu_1 \mu_2 ...\mu_k} A^{*a}_\mu$ and
$D_{\mu_1 \mu_2 ...\mu_k} C^{*a}$
($k = 0, 1, 2 ...$) and where the internal indices are
raised with the Killing metric. Furthermore,
\begin{equation}
H^*(\sigma, {\cal B}) = H^*(\sigma, {\cal D})
\end{equation}
since ${\cal C}$ is contractible.

Now, each $\chi^a_\Delta$ belongs to a copy of the
adjoint representation of
the Lie algebra ${\cal G}$. The algebra
${\cal V}$ of polynomials in
$\chi^a_\Delta$ provides therefore a
representation $\rho : {\cal G}
\rightarrow {\cal V}$ of ${\cal G}$.  Since
${\cal G}$ is semi-simple,
${\cal V}$ splits as a direct sum of
finite-dimensional irreducible
representations,
\begin{equation}
{\cal V} = {\cal V}_0 \oplus (\oplus_{k>0} {\cal V}_k), \;
{\cal V}_0 = \oplus_{\alpha} {\cal V}_{0J\alpha}
\end{equation}
(the space of polynomials of degree $\leq n$
with derivatives up to order
$k$ is invariant and finite-dimensional
for arbitrary $n$ and $k$).  In (21),
${\cal V}_0$ stands for the sum of the
one-dimensional trivial
representations ${\cal V}_{0 \alpha}$,
i.e., contains all the invariant polynomials in
$\chi^a_\Delta$,
while ${\cal V}_k$ denotes the irreducible
non-trivial representations.  Note
that the trivial representation occurs an
infinite number of times (one can form
an infinite number of linearly independent
invariant polynomials in the $\chi^a_\Delta$), so that
${\cal V}_0$ is infinite-dimensional.

The differential $\sigma$ is nothing
but the coboundary operator for the cohomology
$H^*({\cal G, V})$ of the
Lie algebra ${\cal G}$ in the representation
${\cal V}$.  According
to Whitehead theorem, only the invariant
subspace of the trivial
representation contributes to the cohomology and so
\begin{equation}
H^*(\sigma) = H^*({\cal G, V}_0).
\end{equation}
As it follows from the work of \cite{Chev},
the cohomology
$H^*({\cal G, V}_0)$ is the tensor product
of  ${\cal V}_0$ by
the cohomology $H^*({\cal G})$ of the
Lie algebra ${\cal G}$.
$H^*({\cal G})$ is isomorphic to the
algebra of the invariant
cocycles on ${\cal G}$.  Both ${\cal V}_0$
and $H^*({\cal G})$
are free graded-commutative algebras, and
the most general
element of $H^*(\sigma) = H^*({\cal G, V}_0)$
can be taken to have as
representative
\begin{equation}
\sum_{i,J} \alpha_{iJ} P^i(\chi^a_\Delta) E^J(C^a)
\end{equation}
where (i) $P^i(\chi^a_\Delta)$ are a set of
independent generators of
${\cal V}_0$; and (ii) $E^J(C^a)$ are a set of
independent generators of
$H^*({\cal G})$, which are known to be in number
equal to the rank of ${\cal G}$ (``primitive forms").
The independence
of the generators in cohomology means that if
$\sum_{i,J} \alpha_{iJ} P^i(\chi^a_\Delta) E^J(C^a) =
\sigma$(something),
then $\alpha_{iJ} = 0$ for all $i,J$.  That is, if
 an invariant polynomial of the
type (23) is $\sigma$-exact, then it is identically zero.

\section{Cohomology of $\delta$ in ${\cal V}_0$}

 Let $P(\chi^a_\Delta)$ be an invariant polynomial
that is $\delta$-closed and of positive antighost
number,
\begin{equation}
\delta P = 0, \; antigh P > 0, \; P \in {\cal V}_0 .
\end{equation}
By (17), one has
\begin{equation}
P = \delta T
\end{equation}
but there is a priori no guarantee that $T$
belongs also to ${\cal V}_0$.

\noindent
{\bf Theorem 2} : {\it One may choose $T$
in (25) to be in ${\cal V}_0$.}

\noindent
{\bf Proof} : the linear operator $\delta$
is defined in the space ${\cal V}$ of the
polynomials in $\chi^a_\Delta$ and commutes
with the representation $\rho$
of ${\cal G}$ in ${\cal V}$,
\begin{equation}
\delta \rho = \rho \delta.
\end{equation}
Hence it maps the invariant subspaces
${\cal V}_k$ of (21) on invariant
subspaces.  Since ${\cal V}_k$ is
irreducible for $k \not= 0$, the subspace
$\delta {\cal V}_k$ is either isomorphic
to ${\cal V}_k$ and yields an
equivalent representation, or is equal to
$0$.  This implies that
$\delta X $ has no component along the
subspace ${\cal V}_0$ of the trivial
representation if $X \in {\cal V}_k, k \not= 0$,
 i.e. $\delta {\cal V}_k \cap
{\cal V}_0 = 0$ for $k \not= 0$.  Let
$T = T_0 + T_1 + T_2 + ...$
be the decomposition of $T$ along the
subspaces of (21).  Since $\delta T = P$
belongs to ${\cal V}_0$, it follows
that $\delta (\sum_{k>0} T_k) = 0$, so
that one may choose $T = T_0 \in {\cal V}_0$
in (25).  This proves the
theorem.

\section{Cohomology of $s$}

We can now prove the theorem

\noindent
{\bf Theorem 3} : {\it Let $P$ be a polynomial
in the algebra ${\cal A}$ which is
(i) BRST-invariant; and (ii) of ghost
number zero.  Then}
\begin{equation}
P =  M + s Q
\end{equation}
{\it where $M$ is a gauge invariant polynomial
in the field
strengths $F^a_{\mu \nu}$
and their derivatives up to a finite order,
and where $Q \in {\cal A}$.}

\noindent
{\bf Proof} : From Theorem 1, one has
$P = R + sQ'$ where $R \in {\cal B}$
and $sR = 0$.  Let us decompose $R$
according to the antighost number,
\begin{equation}
R = R_0 + R_1 + R_2 + ...+ R_L, \;\;
 antighR_k = k,\;\; k=0,1,...L.
\end{equation}
In virtue of Eqs (1) and (16), the
equation $sR = 0$
is equivalent to the chain of equations
\begin{equation}
\sigma R_0 + \delta R_1 = 0, \; ... ,\;
\sigma R_{L-1} + \delta R_L = 0, \;
\sigma R_L = 0.
\end{equation}
If $L=0$, then $R=R_0$ does not contain
the antifields.  Since its ghost
number vanishes, it does not contain the ghosts either.
Our analysis of the $\sigma$-cohomology
implies then that $R_0$ is an invariant
polynomial in the field strengths and
their derivatives ($\sigma R_0 = 0$),
which establishes the theorem. So let us
assume that $L>0$.
{}From $\sigma R_L = 0$, one gets, as
explained above
\begin{equation}
R_L = \sum_J P^J(\chi^a_\Delta) E^J(C^a) + \sigma Q_L,
\end{equation}
where $P^J \in {\cal V}_0$ and $E^J(C^a)$
are the primitive forms.
By a redefinition of $R_{L-1}$ if necessary,
one can absorb $Q_L$ in a
$s$-variation.  So, let us take $Q_L = 0$.
The $\delta$-variation
of $R_L$ is an invariant polynomial of the
type (23),
$\delta R_L = \sum_{J}\delta P^J(\chi^a_\Delta) E^J(C^a) $
which, by (29), must
be $\sigma$-exact.  By our analysis of
the cohomology of $\sigma$, it must
thus vanish,
\begin{equation}
\delta R_L = 0,
\end{equation}
i.e., $\delta P^J = 0$.  By Theorem 2, this
implies $P^J = \delta
T^J$ with $T^J \in {\cal V}_0$ and thus
\begin{equation}
R_L = \delta T_{L+1},\;\; T_{L+1} =
 \sum_J T^J(\chi^a_\Delta) E^J(C^a),
\end{equation}
with $\sigma T_{L+1} = 0$. Accordingly,
\begin{equation}
R_L = s T_{L+1}
\end{equation}
and one can remove $R_L$ from $R$ by adding a
$s$-variation.  Once this is
done, $R$ contains only components of antighost
numbers $0,1 ...$ up to
$L-1$.  Going on in the same fashion, one then
removes successively $R_{L-1},
R_{L-2} ...$ up to $R_1$.  This yields
\begin{equation}
R = R'_0 + s T'
\end{equation}
where $R'_0$ is an invariant polynomial in the
field strengths and their
derivatives.  Thus one gets
\begin{equation}
P = M + s Q
\end{equation}
with $M = R'_0$ and $Q = Q' + T'$.  This proves the theorem.

\vfill
\eject

\section{Comments}

(1) The above analysis proves that in each
equivalence class of
$H^0(s)$, one can find a representative
that does not
involve the ghosts or the antifields and
that is strictly
gauge invariant.  This
representative, of course, is not unique since
there exist invariant
polynomials in the field strengths and their
derivatives which vanish when the
equations of motion $D_\mu F^{a \mu \nu} = 0$
hold, and which can thus be
written as $s$-variations.  The theorem 3
strengthens the results of
\cite{Fisch,Henn2} in that
(i) the work of \cite{Fisch,Henn2} indicates
that the local BRST
cohomology in degree zero is isomorphic
with the set of on-shell
gauge invariant local polynomials in the
field variables and their
derivatives for a general gauge theory (with
the identification
of two such polynomials that coincide on-shell),
but does not guarantee that one can
find strongly gauge invariant polynomials in
each equivalence class;
(ii) the work of \cite{Fisch,Henn2} shows that
the equations
$s U = 0, antigh U > 0$ imply $U = s V$ for some
$V$ but does not guarantee,
as done in this letter,
that $V$ has a finite expansion in the antighost
number ($V$ could a priori
contain an infinite number of terms of
arbitrarily high antighost number, and
so, not be a polynomial).

Although it is the strong version expressed
by Theorem 3 that has been
invoked in renormalization theory, it appears
that the isomorphism of
$H^0(s)$ with the set of on-shell gauge
invariant polynomials is just
sufficient on physical grounds.  Indeed,
the physical matrix elements
of BRST-exact operators (with s given by
the sum (1)) vanish and so,
these operators are physically irrelevant
\cite{Henn1,HT}.
Accordingly, provided
the BRST symmetry is not anomalous so that
BRST invariant operators mix
only with BRST invariant operators, and BRST
exact operators mix only
with BRST exact ones, then, the mixing is
well defined in cohomology.
By the isomorphism mentioned above, this
means that only gauge-invariant
operators are physically relevant. The isomorphism of
$H^*(s)$ with the set of gauge invariant
functionals admits an
extension to non-local operators
\cite{Fisch,Henn1,HT}.

(2) Theorem 3 can be extended to other
values of the ghost number.  One
shows along the same lines the existence,
in each equivalence
class of $H^*(s)$, of a representative
annihilated by $\sigma$.

(3) Finally, our analysis does not cope
with the more complicated
question of
computing $H(s = \delta + \sigma \mid d)$.  The cohomology of
$H(\sigma \mid d)$ has been the subject
of various works (\cite{Dixon2,DTV,Band,Brandt,DubV}
and references therein).
It would be of interest to extend those
results to $H(s \mid d)$.  We plan
to return to this question elsewhere.

\section{Acknowledgements}

The author is grateful to John Collins
for a question that prompted this
work and to Glenn Barnich, Michel
Dubois-Violette, Jim Stasheff,
Michel Talon, Claudio Teitelboim and
Claude Viallet for useful conversations.
This work has been
supported in part by research funds from
F.N.R.S. and a research contract with
the Commission of the European Communities.

\end{document}